\documentclass[12pt, a4paper]{article}
\usepackage{jheppub}
\usepackage{physics}
\usepackage{amsfonts}
\usepackage{amssymb}
\usepackage{amsthm}
\usepackage{bm}
\usepackage{graphicx}
\usepackage{url}
\usepackage{lmodern}
\usepackage{subfigure}
\usepackage[section]{placeins}
\usepackage{xcolor}
\usepackage{float}
\usepackage{multirow}
\usepackage[utf8]{inputenc}
\usepackage[T1]{fontenc}
\definecolor{refkey}{gray}{0.45}
\definecolor{labelkey}{RGB}{155,48,48}
\hypersetup{
  colorlinks=true,
  citecolor=magenta,
  linkcolor=blue,
  urlcolor=violet
 }
 \usepackage{tikz}
\usetikzlibrary{decorations.pathmorphing}

\usepackage{chngcntr}
\counterwithout{equation}{section}

\newcommand{\del}{\partial}
\newcommand{\mi}{\mathrm{i}}
\def\be{\begin{equation}}\def\ee{\end{equation}}
\newcommand{\baa}{\begin{equation}\begin{aligned}}
\newcommand{\ea}{\end{aligned}\end{equation}}

\title{Strong Cosmic Censorship in Two Dimensions}

\preprint{TIFR/TH/20-42}

\author{Upamanyu Moitra}

\affiliation{Department of Theoretical Physics, Tata Institute of Fundamental Research \\
1 Homi Bhabha Road, Colaba, Mumbai -- 400005, India}

\emailAdd{upamanyu@theory.tifr.res.in}

\abstract{We study the behaviour of a free massive scalar wave-packet near the Cauchy horizon of an $\mathrm{AdS}_2$ black hole and find that it becomes infinitely differentiable for smooth initial data, independently of the parameters describing the spacetime or the scalar.  This indicates a violation of the strong cosmic censorship conjecture at the classical level. We discuss our result in connection with some recent observations of violation of the conjecture for certain nearly extremal black holes. }

\dedicated{\centering \vspace{1in} Dedicated to\\my grandparents\\
Dhiranando and Tripti Roy}

\makeatletter
\gdef\@fpheader{}
\makeatother
\begin{document}
\maketitle

\section{Introduction}

The general theory of relativity is arguably the most elegant classical theory of physics. The presence of Cauchy horizons in some solutions of the Einstein equations (e.g., the  Reissner-Nordstr\"{o}m and Kerr solutions), however, is an unwelcome feature of general relativity, since one loses the predictive power of general relativity beyond this horizon.  One needs to introduce additional elements to reinstate predictivity. The strong cosmic censorship conjecture \cite{Penrose:1900mp} is one such element, which asserts that Cauchy horizons are generically unstable.

The question of the stability of Cauchy horizons has been studied for a long time now (see e.g., \cite{Mcnamara:1978, Chandra:1982, Poisson:1989zz,  Ori:1991zz, Brady:1998au, Dafermos:2003wr}) and the strong cosmic censorship conjecture is believed to be true for a wide class of black holes. It has received some attention in the recent years --- in particular, it has been observed that strong cosmic censorship is violated for certain black holes (e.g., the BTZ black hole \cite{Dias:2019ery} and charged black holes in de Sitter spacetime \cite{Dias:2018etb, Dias:2018ufh, Cardoso:2017soq}) when the black holes are sufficiently near extremality. A wide class of extremal black holes have a common aspect --- a near-horizon $\mathrm{AdS}_2$ factor \cite{Kunduri:2007vf}. Even when the black holes are nearly extremal, an appropriate scaling does give rise to a near-horizon $\mathrm{AdS}_2$ black hole geometry (see e.g., \cite{Moitra:2019bub}).  This naturally begs the question whether the violation has anything to do with the $\mathrm{AdS}_2$ factor. This question is the motivation for this work and we will answer this question in the affirmative.

Gravity in two spacetime dimensions is rather special  -- the Einstein-Hilbert action is a topological invariant and cannot describe dynamics. One needs to couple gravity to other fields, typically scalars, to get interesting features. One such model is Jackiw-Teitelboim (JT) gravity \cite{Teitelboim:1983ux, Jackiw:1984je}, which describes a gravity-dilaton system. The JT model has been extensively studied \cite{Maldacena:2016upp} in the recent years --- its connection with the quantum mechanical Sachdev-Ye-Kitaev model \cite{Sachdev:1992fk, Kitaev, Maldacena:2016hyu} and its variants has attracted considerable attention from the community. It has been successful in describing features of nearly extremal black holes with a near-horizon $\mathrm{AdS}_2$ geometry (see \cite{Nayak:2018qej, Moitra:2018jqs, Moitra:2019bub, Moitra:2019xoj} and references therein).

We shall be concentrating on an $\mathrm{AdS}_2$ black hole with a Cauchy horizon. Our strategy would be to consider a minimally coupled massive probe scalar and examine its behaviour near the Cauchy horizon. We find that the scalar field is infinitely differentiable ($C^\infty$) for initial smooth data on appropriate hypersurfaces.  Our result is stronger than the case for BTZ black holes. The authors of \cite{Dias:2019ery} found that a neutral scalar would be $C^k$ at the Cauchy horizon, with $k$ determined by the parameters of the geometry and the conformal dimension of the scalar field. It was found that $k$ could be made arbitrarily large by making the black hole approach extremality. In our case, the violation of strong cosmic censorship does not depend on the conformal dimension or the parameters describing the geometry (this fact is actually obvious from the point of view of symmetry).

Our analysis will be very similar to that in \cite{Dias:2019ery} and therefore, we will be brief in describing the calculation, drawing attention to the most important points, especially those that differ from \cite{Dias:2019ery}. Readers concerned about various subtle aspects not explicitly commented on here would find the details in   \cite{Dias:2019ery} illuminating. The question of strong cosmic censorship in BTZ black holes was addressed in \cite{Balasubramanian:2004zu} and more recently in, e.g., \cite{Papadodimas:2019msp, Balasubramanian:2019qwk, Emparan:2020rnp,Bhattacharjee:2020gbo}. 
\section{Strong Cosmic Censorship in \boldmath $\mathrm{AdS}_2$}
We will consider a minimally coupled massive scalar field in the background of an $\mathrm{AdS}_2$ black hole.  The line element can be expressed as
\baa
\label{metgauge}
\dd s^2 &= -  (r^2 - r_0^2) \dd t^2 + \frac{\dd r^2}{r^2 - r_0^2}.
\ea
There is a linearly varying time-independent dilaton as well (in addition to possibly other 2d fields), but that does not play an important role in our considerations\footnote{Actually, purely within the JT model, the Cauchy horizon lies at a region where the dilaton (which couples to the Ricci scalar) turns negative. This is, however, easy to fix --  the presence of the toplogical Einstein-Hilbert term with a suitable coefficient could ensure that the coefficient of the Ricci scalar is positive even inside the Cauchy horizon.  One can also achieve this by adding other fields. See,  e.g.,  \cite{Castro:2008ms}.}. We have set $L_{\mathrm{AdS}_2}=1$.

The event horizon is located at $r=r_0$ and the Cauchy horizon is at $r=-r_0$. It is an important property of $\mathrm{AdS}_2$ black holes, not shared by other black holes even in two dimensions,  that the surface  gravity at the inner and outer horizon are equal:
$$
\kappa_\pm = r_0.
$$
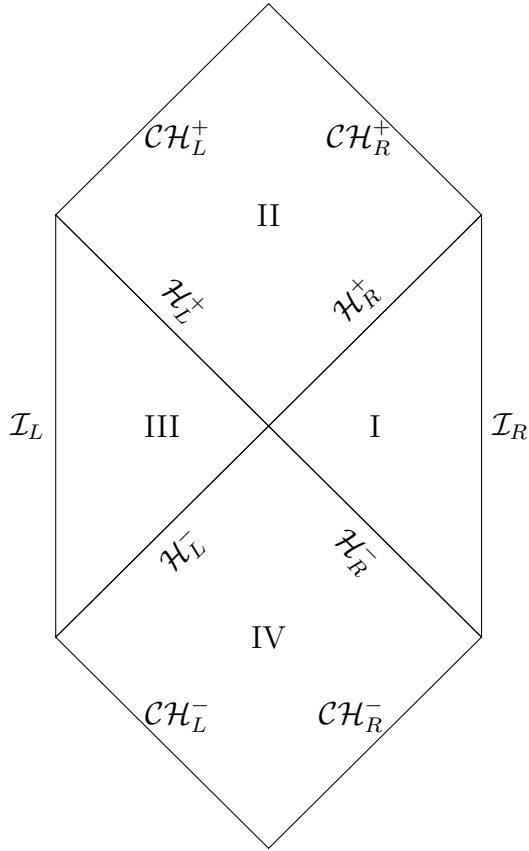
\begin{figure}
\centering
\begin{tikzpicture}[scale=0.70]
\node (I)    at ( 2,0)   {I};
\node (III)   at (-2,0)   {III};
\node (II)  at (0, 4) {II};
\node (IV)   at (0,-4) {IV};
\path 
  (I) +(2,4)  coordinate (rItop)
       +(-2,0) coordinate (rIbot)
       +(2,-4)   coordinate  (rIright);
\draw  (rItop) --  (rIbot) -- (rIright) -- node[right]   {${\cal I}_R$}   (rItop)  --  cycle;
\path
  (II) +(0,-4)  coordinate (urIbot)
       +(4,0) coordinate (urIright)
       +(0,4)   coordinate (urItop)
       +(-4,0) coordinate (urIleft);
 \draw  (urIbot) -- node[midway, above, sloped]    {${\cal H}^+_R$} (urIright) -- node[midway, below]    {${\cal CH}^+_R \quad$} (urItop) -- node[midway, below]    {$\quad  {\cal CH}^+_L  $} (urIleft)  -- node[midway, above, sloped]    {${\cal H}^+_L$}  cycle;
 \path 
  (III) +(-2,4)  coordinate (lItop)
       +(2,0) coordinate (lIbot)
       +(-2,-4)   coordinate  (lIright);
      \draw  (lItop) -- (lIbot) -- (lIright) -- node[left]   {${\cal I}_L$}  (lItop)  -- cycle;
 \path
  (IV) +(0,-4)  coordinate (utIbot)
       +(4,0) coordinate (utIright)
       +(0,4)   coordinate (utItop)
       +(-4,0) coordinate (utIleft);
\draw  (utIbot) -- node[midway, above]    {${\cal CH}^-_R\quad \,\,\,$}  (utIright) -- node[midway, below, sloped]    {${\cal H}^-_R$}  (utItop) -- node[midway, below, sloped]    {${\cal H}^-_L$}   (utIleft)  -- node[midway, above]    {$\quad{\cal CH}^-_L$}  cycle;
\end{tikzpicture}
\caption{The Penrose diagram for the eternal $\mathrm{AdS}_2$ black hole: In this diagram, $\cal{CH}$ refers to  the Cauchy horizons , $\cal{H}$ refers to the event horizons and ${\cal I}$ refers to the asymptotic boundaries, while $+(-)$ refers to future (past). This diagram,  along with lines representing the timelike singularity,  repeats infinitely in the vertical direction for the maximal analytic extension. In this paper, we are interested in regions I and II. }
\label{fig1:pendia}
\end{figure}
We can set $r_0=1$ in the metric by a simultaneous rescaling of $t$ and $r$,
$$
t \to   \frac{t}{r_0} ,\quad r \to r r_0,
$$
so that the metric  takes the form,
\baa
\label{rescmet}
\dd s^2 &= -  (r^2 -1) \dd t^2 + \frac{\dd r^2}{r^2 - 1}.
\ea
Note that when there is a varying dilaton present, as in our case,  such a rescaling does change the dilaton. However, for the discussion below involving a scalar, which couples only to the metric, this issue is not important. The asymptotic timelike boundary is located at a value of $r=r_b \gg 1$. For our purpose, we can essentially take $r_b \to \infty$. The causal structure of the spacetime is depicted in Figure \ref{fig1:pendia}.

It is our ability to scale $r_0$ out of the problem that makes the question of cosmic censorship independent of $r_0$. Whether cosmic censorship is violated or not is usually determined by the quantity $$\beta \equiv \frac{\alpha_r}{\kappa_-}, $$
where $\alpha_r$, known as the \emph{spectral gap}, is the negative imaginary part of a quasi-normal mode. Since the frequency is necessarily proportional to $r_0$ (as is $\kappa_-$) because of the scaling symmetry, it follows that $\beta$ is independent of $r_0$.

The scalar equation reads,
\baa
\label{kgeq}
{\nabla}^2 \psi - \mu^2 \psi = 0,
\ea
where $\mu$ is the mass of the scalar. We seek a mode solution of the form,
$$
\psi(\omega; t,r) = e^{ - \mi \omega t} \psi ( r). 
$$
Note that, we use the same letter $\psi$ whether or not $t$ is part of the argument. What we are referring to will be obvious from the context.  
Eq.\eqref{kgeq} reads,
\baa
\label{ckgfull}
\omega^2 \psi  + (r^2 -1) \del_r (  (r^2-1)  \del_r \psi) - \mu^2 (r^2 -1) \psi =0.
\ea
It is easy to solve this equation in terms of hypergeometric functions \cite{AShandbook}. We introduce a new variable, 
\baa
\label{defvarx}
y = \frac12 (r+1).
\ea
This maps the Cauchy horizon, the event horizon and the asymptotic boundary to $y=0$, $y=1$ and $y=\infty$, the regular singular points of the hypergeometric differential equation. There exists a basis of a pair of linearly independent solutions to eq.\eqref{ckgfull} about each of these singular points. It is worth noting that even in case of the BTZ black hole, the solution can be expressed in terms of hypergeometric functions \cite{Balasubramanian:2004zu, Dias:2019ery}. We can write down the solutions about these points as follows:
\begin{itemize}
\item About the Cauchy horizon:
\baa
\label{solchbasis}
\psi^{\cal CH}_{\rm out} (y) &= y^{ \frac12 (c-1) } |1-y|^{ \frac12 (a+b-c) } \,_2 F_1 (a,b; \,  c;\, y),  \\
\psi^{\cal CH}_{\rm in}  (y) &= y^{- \frac12 (c-1) } |1-y|^{ \frac12 (a+b-c) } \,_2 F_1 (1+a-c,1+b-c; \,  2-c;\, y).
\ea

\item About the event horizon:
\baa
\label{solehbasis}
\psi^{\cal H}_{\rm out} (y) &= y^{ \frac12 (c-1) } |1-y|^{ -\frac12 (a+b-c) } \,_2 F_1 (c-a,c-b; \, 1+ c-a-b;\,1- y) , \\
\psi^{\cal H}_{\rm in} (y) &=y^{ \frac12 (c-1) } |1-y|^{ \frac12 (a+b-c) } \,_2 F_1 (a,b; \, 1+a+b-c;\,1- y) .
\ea
\item About the asymptotic boundary:
\baa
\label{solinfbasis}
\psi^{\cal I}_{\rm nn} (y) &= y^{ \frac12 (c-1-2a) } |y-1|^{ \frac12 (a+b-c) } \,_2 F_1 \pqty{ a, 1+a-c; \, 1+ a-b;\, \frac{1}{y}}, \\
\psi^{\cal I}_{\rm nor} (y) &=y^{ \frac12 (c-1-2b) } |y-1|^{ \frac12 (a+b-c) } \,_2 F_1 \pqty{ b, 1+b-c; \, 1+ b-a;\, \frac{1}{y}}.
\ea
\end{itemize}

Here, the parameters of the hypergeometric function are given by,
\baa
\label{hgpara}
a &= \Delta_-  - \mi  \omega, \\
b &= \Delta_+ - \mi  \omega, \\
c &= -\mi \omega  + 1,
\ea
and,
\baa
\label{sddefs}
\Delta_\pm = \frac12 \pqty{ 1 \pm \sqrt{1+ 4 \mu^2} } \equiv \frac12 \pqty{ 1 \pm \nu }.
\ea

Let us make a few comments here. As suggested by the names of the functions, the solutions are ingoing or outgoing  on the appropriate part of the various horizons, as can be seen by going to the in/outgoing Eddington-Finkelstein coordinates\footnote{The ingoing coordinate is $v= t+r^*$ and the outgoing coordinate is $u=t-r^*$, with $r^* = \int \dd r (r^2 -1)^{-1} = \frac12 \log ( |r-1|/|r+1|)$.}. For example, $\psi^{\cal CH}_{\rm out}$ is smooth ($\sim e^{-i \omega u}$) on ${\cal CH}^+_R$, but not on ${\cal CH}^+_L$, and  $\psi^{\cal CH}_{\rm in}$ is smooth $( \sim  e^{-i \omega v})$ on ${\cal CH}^+_L$, but not on ${\cal CH}^+_R$. The quantities $\Delta_\pm$ describe the asymptotic behaviour of the scalar. We have, for large $y$ (i.e., large $r$),
\baa
\label{psilarge}
\psi^{\cal I}_{\rm nn} &\sim y^{-\Delta_-},  \\
\psi^{\cal I}_{\rm nor} &\sim y^{-\Delta_+}. 
\ea
We assume the scalar satisfies the Breitenl\"{o}hner-Freedman bound \cite{Breitenlohner:1982jf}: i.e., $\nu$ is real: $4 \mu^2  \geq -1$ and so $\Delta_+ \geq  \Delta_-$. The mode  $\psi^{\cal I}_{\rm nn}$ thus describes a non-normalisable mode at the asymptotic boundary, while  $\psi^{\cal I}_{\rm nor}$ describes a normalizable fall-off. (Actually, in a slight abuse of terminology, we continue to call the slower decaying solution non-normalisable even when $0<\nu <1$.)

Using the conventions of \cite{Dias:2019ery}, we can relate the event horizon basis to the boundary basis:
\baa
\label{psiotra}
\psi^{\cal H}_{\rm in}   &= \frac{1}{\cal T} \psi^{\cal I}_{\rm nn} + \frac{\cal R}{\cal T} \psi^{\cal I}_{\rm nor}, \\
\psi^{\cal I}_{\rm nor}   &= \frac{1}{\widetilde{\cal T}} \psi^{\cal H}_{\rm out} + \frac{\widetilde{\cal R}}{\widetilde{\cal T}} \psi^{\cal H}_{\rm in} . 
\ea
Using the well known hypegeometric transformation formul\ae, \cite{AShandbook} we easily obtain,
\baa
\label{ttrcoe}
{\cal T} &= \frac{\Gamma \pqty{ \Delta_+ - \mi \omega) } \Gamma (\Delta_+)  }{\Gamma (1- \mi \omega) \Gamma (\nu) } , \\
{\cal R} &= \frac{\Gamma (-\nu) \Gamma \pqty{ \Delta_+  - \mi \omega } \Gamma (\Delta_+ )  }{\Gamma(\nu) \Gamma \pqty{ \Delta_-  -\mi \omega } \Gamma (\Delta_-  ) } .
\ea
and,
\baa
\label{tilttrcoe}
\widetilde{\cal T} &= \frac{\Gamma \pqty{ \Delta_+ - \mi \omega } \Gamma (\Delta_+  ) }{\Gamma (2\Delta_+) \Gamma (- \mi \omega) } , \\
\widetilde{\cal R} &= \frac{\Gamma (\mi \omega)  \Gamma \pqty{ \Delta_+ - \mi \omega }   }{\Gamma(-\mi \omega)  \Gamma \pqty{ \Delta_+ + \mi \omega   } } .
\ea
The exterior quasi-normal modes (QNMs) can be read off from the poles of ${\cal T}$ or $\widetilde{\cal T}$ and are given by,
\baa
\label{exqnm}
\omega = - \mi (\Delta_++n) \equiv \omega_{\rm ext}  ,
\ea
where $n=0,1,2,\cdots$ (we will continue to refer to all non-negative integers with $n$ throughout this paper).
Note that, in contrast with the results of \cite{Dias:2019ery}, there is only one family of exterior QNMs, which is related to the fact that the surface gravities corresponding to the two horizons are equal.

As emphasised by the authors of \cite{Dias:2019ery}, in exploring the question of strong cosmic censorship, the issue of what they refer to as \emph{interior QNMs} is also relevant. The interior scattering coefficients are defined as follows. The Cauchy horizon basis can be written in terms of the event horizon basis in the following form:
\baa
\label{psiitr}
\psi^{\cal H}_{\rm out} &= {\cal A} \psi^{\cal CH}_{\rm out} +  {\cal B} \psi^{\cal CH}_{\rm in} , \\
\psi^{\cal H}_{\rm in} &= \widetilde{\cal A} \psi^{\cal CH}_{\rm in} + \widetilde{\cal B} \psi^{\cal CH}_{\rm out} .
\ea

An application of the hypergeometric identities again yields,
\baa
\label{abcoe}
 {\cal A} &=  \frac{\Gamma(1+\mi \omega) \Gamma(\mi \omega) }{\Gamma \pqty{ \Delta_+ + \mi \omega } \Gamma \pqty{ \Delta_- + \mi \omega   }  } , \\
  {\cal B} &=  \frac{\Gamma(1+\mi \omega) \Gamma(-\mi \omega) }{\Gamma \pqty{ \Delta_+  } \Gamma \pqty{ \Delta_- } } ,
\ea
and,
\baa
\label{tilabcoe}
\widetilde{\cal A} &=  \frac{\Gamma(1-\mi \omega) \Gamma(-\mi \omega)}{\Gamma \pqty{ \Delta_+ - \mi \omega } \Gamma \pqty{ \Delta_-  - \mi \omega   }  } , \\
  \widetilde{\cal B} &=  \frac{\Gamma(1-\mi \omega) \Gamma(\mi \omega) }{\Gamma \pqty{ \Delta_+ } \Gamma \pqty{ \Delta_-} } .
\ea
Note immediately that in ${\cal B}$ and $\widetilde{\cal B}$, there are no zeroes in $\omega$. Therefore, the ``out-out'' and the ''in-in'' QNMs discussed in the case of BTZ black holes in \cite{Dias:2019ery} do not exist here. There do, however, exist two families of QNM frequencies for each of the ``in-out'' ($ \widetilde{\cal A} =0$) or ``out-in'' ($ {\cal A} =0$) cases. These are given by,
\baa
\label{ominout}
\omega^{(1)}_{\text{in-out}} &= - \mi (\Delta_++n), \\ 
\omega^{(2)}_{\text{in-out}} &=  - \mi (\Delta_-+n),
\ea
and,
\baa
\label{omoutin}
\omega^{(1)}_{\text{out-in}} &= + \mi ( \Delta_+ +n), \\
\omega^{(2)}_{\text{out-in}} &= + \mi ( \Delta_- +n).
\ea
Note that the quantities in \eqref{omoutin} are complex conjugates of \eqref{ominout} above. Further note that $\omega^{(1)}_{\text{in-out}}$ is the same as $\omega_{\rm ext}$, eq.\eqref{exqnm}. Such a coincidence was crucial for establishing the result of \cite{Dias:2019ery}. It will perhaps not surprise the reader that this fact will be crucial for us as well.

\subsection{Classical Wave-Packets}
We are now ready to consider the question of strong cosmic censorship. To do so, one needs to consider appropriate wave-packets on surfaces suitable for an initial and boundary value formulation and examine the behaviour near the Cauchy horizon. The characteristic data could be taken to have support on the future left event horizon (${\cal H}^+_L$), past right event horizon (${\cal H}^-_R$) and right timelike boundary (${\cal I}_R$). The data on these surfaces are described by $X_{{\cal H}^+_L } (\omega)$, $X_{{\cal H}^-_R } (\omega)$ and  $X_{{\cal I}_R } (\omega)$, which correspond to the Fourier transforms of compactly supported data on these surfaces and thus, are entire in $\omega$. We can then carry out the analysis as in \cite{Dias:2019ery} and write the mode solution in region II in the Cauchy horizon basis \eqref{solchbasis}. (We do not reproduce the various subtle arguments extensively covered in reference \cite{Dias:2019ery}.)
\baa
\label{psitot}
\psi (t, r) =  \psi_{\rm out} (t,r) + \psi_{\rm in} (t,r),
\ea
where,
\baa
\label{psisplit}
\psi_{\rm out} (t,r) &= \int \dd \omega \bqty{ X_{{\cal H}^+_L } (\omega) {\cal A} + \pqty{ X_{{\cal H}^-_R } (\omega) \widetilde{\cal R} + X_{{\cal I}_R } (\omega) {\cal T} } \widetilde{\cal B} } \psi^{\cal CH}_{\rm out} (\omega; t, r) ,\\
\psi_{\rm in} (t,r) &= \int \dd \omega \underbrace{  \bqty{X_{{\cal H}^+_L } (\omega) {\cal B} + \pqty{ X_{{\cal H}^-_R } (\omega)  \widetilde{\cal R} + X_{{\cal I}_R } (\omega) {\cal T} } \widetilde{\cal A} } }_{ {\cal G}_{\rm in} (\omega)  } \psi^{\cal CH}_{\rm in} (\omega; t, r).
\ea
We have suppressed the $\omega$-dependence in the scattering coefficients, but they are, of course, $\omega$-dependent, as can be seen from the definitions above.

To address the question of strong cosmic censorship, say, on the right Cauchy horizon\footnote{As  emphasised in \cite{Dias:2019ery}, this is the relevant region for studying strong cosmic censorship for black holes forming out of collapsing matter, as in, say, \cite{Moitra:2019xoj}.} $\mathcal{CH}^+_R$,  we need to consider only $\psi_{\rm in} (t,r) $ which could potentially have non-smooth behaviour there. On the other hand,  $\psi_{\rm out}  (t,r) $ is manifestly smooth here.

Let us now comment on the analytic structure of the different terms of the function ${\cal G}_{\rm in} (\omega)$, which appears in the integrand of $\psi_{\rm in} (t,r)$ in eq.(\ref{psisplit}). 
\begin{itemize}
\item The quantity ${\cal B}$ \eqref{abcoe} has simple poles on the upper half-plane (UHP), $\omega = \mi (n+1)$ and on the lower half-plane (LHP), $\omega = - \mi n$. This includes a pole on the real axis, $\omega = 0$ for $n=0$.
\item  $\widetilde{\cal A}$ \eqref{tilabcoe} has a simple pole at $\omega =0$ and \emph{double} poles at  $\omega = -\mi (n+1)$ -- note that ${\cal B}$ has simple poles at the location of these double poles. The zeroes of $\widetilde{\cal A}$ \eqref{tilabcoe} are given by eq.\eqref{ominout}, namely $\omega = \omega_{\rm ext}$ and $\omega = - \mi (\Delta_- +n)$.
\item  $\widetilde{\cal R}$ \eqref{tilttrcoe} has simple poles at $\omega = \mi (n+1)$ and $\omega = \omega_{\rm ext}$ \eqref{exqnm}. It has zeroes at $\omega = -\mi (n+1)$ and at $\omega = + \mi (\Delta_+ +n)$.
\item Finally, ${\cal T}$ \eqref{ttrcoe} has simple poles at $\omega = \omega_{\rm ext}$ \eqref{exqnm} and zeroes at $\omega = - \mi (n+1)$.
\end{itemize}

There is a simple pole of the integrand of $\psi_{\rm in} (t,r)$ \eqref{psisplit} on the real axis at $\omega = 0$, due to those in $\widetilde{\cal A}$ and ${\cal B}$. A careful argument in \cite{Dias:2019ery}, however, shows that the pole is artefact of the basis used to construct the solution and the integration contour on the real axis can be deformed into the LHP to avoid this pole. All we need to do now is find the poles of ${\cal G}_{\rm in} (\omega)$ in the LHP.

It turns out that ${\cal G}_{\rm in} (\omega)$ in analytic on the LHP, except for only \emph{simple} poles at $\omega = -\mi (n+1)$:
\begin{itemize}
\item From the discussion above, it is obvious that ${\cal B}$ has simple poles on the LHP at $\omega = - \mi (n+1)$.
\item  Now, note that $\widetilde{\cal R}$ and $\cal T$ both have simple poles at $\omega = \omega_{\rm ex}$, but this pole is removed by the zeroes of $\widetilde{\cal A}$ (which multiplies both in $\psi_{\rm in} (t,r)$ \eqref{psisplit}) at the same location. As a result, there are no poles of $ {\cal G}_{\rm in} (\omega)$ at $\omega = \omega_{\rm ex}$.
\item Finally, note that $\widetilde{\cal A}$ has double poles at $\omega = -\mi (n+1)$. However, both $\widetilde{\cal R}$ and $\cal T$ have zeroes at the same locations. Since $\widetilde{\cal A}$ is multiplied by either $\widetilde{\cal R}$ or $\cal T$, the potential double poles of $ {\cal G}_{\rm in} (\omega)$ are  turned into \emph{simple} poles.
\end{itemize}

We now examine the behaviour of the mode function near the Cauchy horizon at $y=0$. In the outgoing Eddington-Finkelstein coordinate ($u= t - r^*$), we have,
$$
\psi_{\rm in} (\omega; t,r) \sim e^{ - \mi \omega u } y^{ \mi  \omega}.
$$
Since the only actual simple poles are at $\omega = - \mi (n+1)$, it follows immediately that the contribution of such a pole gives rise to smooth behaviour in $y$:
\baa
\psi _{\rm in} \sim e^{-(n+1)u } y^{n+1}.
\ea

Clearly, therefore, the right component of the scalar is smoothly vanishing on the right Cauchy horizon ($y \to 0$, $u $ varying)\footnote{The bifurcate Cauchy horizon ($u\to-\infty$) is not too important for our considerations; see \cite{Dias:2019ery}.}. The left component is smooth anyway. The scalar $\psi$ is thus $C^\infty$ on the Cauchy horizon and this indicates a classical violation of the strong cosmic censorship conjecture.

\section{Discussion}

The violation of strong cosmic censorship in ${\rm AdS}_2$ is dependent on no fewer than two coincidences. One coincidence was also present for BTZ black holes \cite{Dias:2019ery}, an interior quasi-normal frequency being equal to an exterior quasi-normal frequency. This fact was responsible for the eventual violation of strong cosmic censorship in BTZ black holes.  In our case, an additional coincidence --- the equality of surface gravity on the event and inner horizons --- ensured that the violation was stronger.

It is worth remarking that the violations of strong cosmic censorship are observed,  both in this paper and in \cite{Dias:2019ery},  in rather special low-dimensional geometries.  The black holes in these scenarios,  in contrast with their counterparts in higher dimensions,  can actually be obtained as coordinate transformations of a global $\mathrm{AdS}$ spacetime.  The BTZ black hole is a quotient of global $\mathrm{AdS}_3$ and the black hole geometry considered here is a patch of global $\mathrm{AdS}_2$ (it is worth emphasising that globally the spacetime is \emph{not} global $\mathrm{AdS}_2$,  but there is a genuine black hole here because of a nontrivial dilaton background).  From this perspective,  the violation can be thought of as a consequence of the enhanced symmetry of these black hole spacetimes in lower dimensions,  as the horizons are somewhat less distinguished because of the symmetry. 

We considered a classical scalar wave-packet. Classical fields of other kinds could be considered (see \cite{Dias:2019ery}), but since the equations of motion are eventually expected to be the same in form as the scalar equation, the results are expected to be no different. It is worth emphasising that the violation is independent of the mass of the scalar field. As long as there is no instability associated with the violation of the Breitenl\"{o}hner-Freedman bound, all fields are expected to violate strong cosmic censorship.

One interesting aspect to consider would be a calculation of holographic correlators in the Lorentzian signature \cite{Balasubramanian:1998de}. A calculation similar to one in \citep{Balasubramanian:2004zu} suggests that the holographic one-point function does not receive a divergent contribution from the Cauchy horizon. There has been a recent work \cite{Grinberg:2020fdj} which explores the issue of probing behind the Cauchy horizon.  It would be interesting to probe the deep interior of a black hole from a quantum theory like the SYK model.

While $\mathrm{AdS}_2$ geometries arise from near-extremal black holes, it is clear that not all such higher dimensional geometries would show the violation of strong cosmic censorship, since the boundary conditions on the $\mathrm{AdS}_2$ boundary connecting to the UV geometry in such cases, different from the one considered here,  would enforce strong cosmic censorship, because the analytic structure of the function ${\cal G}_{\rm in} (\omega)$ would be dramatically altered.  See \cite{Dias:2019ery} for some related comments.

What our work \emph{does} show is that strong cosmic censorship is generically violated for $\mathrm{AdS}_2$ black holes, when we have the right boundary conditions. This is a pleasing result because it is consistent with the observed violation in the case of some near-extremal black holes in rather different situations (e.g., the BTZ black hole or RN-dS black hole), as mentioned in the introduction -- and this work connects these diverse situations with the same thread.   One might wonder why the violation is stronger ($C^\infty$) in our case. This has to do with the fact that the minimally coupled scalar in higher dimensions is not the same as a minimally coupled scalar in two dimensions (in our case,  the scalar is insensitive to the breaking of isometries of the $\mathrm{AdS}_2$ spacetime by the varying dilaton -- see also the comment on symmetries made above).  A dimensionally reduced minimally coupled scalar will necessarily have a coupling with the dilaton of two-dimensional gravity ($\Phi^d$: the volume of the transverse space);
\baa
S_\psi \propto - \int \dd[2] x \sqrt{-g} \Phi^d \pqty{ (\nabla \psi)^2 + \mu^2 \psi^2 }.
\ea
This would certainly change the behaviour of the scalar near the Cauchy horizon in some respects. Furthermore, the corrections to the $\mathrm{AdS}_2$  geometry arising from non-linear dilaton terms of the dimensionally reduced action  would also make the surface gravity on the two horizons different, which would change our conclusions.

It would be interesting to consider other aspects of strong cosmic censorship in 2d, e.g., effects of \emph{quantum} fields. We leave such studies for the future.

\acknowledgments 
I would like to thank Harvey Reall for his nice lectures on strong cosmic censorship at TIFR in September 2019.  My research is supported by the Department of Atomic Energy, Government of India under Project No. 12-R{\&}D-TFR-5.02-0200 and the Infosys Endowment for the Study
of the Quantum Structure of Spacetime. I also register my gratitude to the people of India for their steady support for the study of basic sciences.

\bibliographystyle{JHEP}

\bibliography{sccref}
\end{document}